\documentclass{aa}

\usepackage[switch]{lineno}

\usepackage{subfig}
\usepackage{soul}	
\usepackage{xcolor}
\usepackage[colorlinks=true, allcolors=blue]{hyperref}

\usepackage{graphicx}
\usepackage{txfonts}
\begin{document} 

 \titlerunning{The gamma-ray emission of SNR G106.3+2.7}
 \authorrunning{P. Cristofari et al.}
\title{Can a single supernova remnant account for the gamma-ray emission of G106.3+2.7?}
   \author{P. Cristofari,
          \inst{1}\thanks{pierre.cristofari@obspm.fr}
            G. Emery,\inst{2, 3}
            T. Lubrano di Vavaria,\inst{2}
            H. Costantini,\inst{2}
            F. Cassol,\inst{2}
           M.-S. Carrasco  \inst{2}
           \and
           B. Le Nagat Neher\inst{1}
          }

   \institute{
   \inst{1} Laboratoire d’étude de l’Univers et des phénomènes eXtrêmes, LUX, Observatoire de Paris, Université PSL, Sorbonne Université, CNRS, 92190 Meudon, France\\
    \inst{2} Aix Marseille Univ, CNRS/IN2P3, CPPM, Marseille, France \\
    \inst{3} Instituto de Astrofísica de Andalucía, IAA-CSIC, Glorieta de la Astronomía s/n, 18008 Granada, Spain
             }

   \date{}

 
  \abstract
   {SNR G106.3+2.7 is a complex TeV emitting source whose emission is still poorly understood. It has especially been at the center of numerous discussions on its potential for being a supernova remnant (SNR) PeVatron, since its gamma-ray spectra seems not to exhibit any significant suppression in the multi--TeV range, up to $\sim 600$ TeV, thereby indicating the presence of $\sim$ PeV particles. 
}
   { We study the hypothesis in which a SNR evolving in a clumpy or cloudy environment is powering the TeV gamma-ray emission, detected mainly from two regions, the "head" and the "tail". We discuss the implications of such an hypothesis. 
}
   {We rely on a simple physically motivated analytical modeling of the shock dynamics and of the content of accelerated particles and confront it to available gamma-ray observations. 
}
   {We find that the current observed TeV gamma-ray emission in the head and tail regions can be accounted for by an active single SNR, with a natural hardening of the spectrum due to the expansion in a clumpy medium or escaping to a dense region in the tail. However, in all scenarios, the broadband gamma-ray emission from the GeV range to the $\gtrsim 100$ TeV range is difficult to reconcile with a standard SNR - whether originating from a thermonuclear or a core-collapse supernova - and instead points toward an association with the pulsar.}
   {}


   \maketitle
%

\section{Introduction}
Supernova remnants (SNRs), as non-relativistic collisionless shock waves, are known to accelerate charged particles up to the very--high--energy (VHE) domain. 
The main mechanism responsible for the acceleration of particles is the first-order Fermi mechanism called "Diffusive shock acceleration" (DSA). Although the general picture of DSA is well depicted~\citep{axford1977,krymskii1977,blandford1978,bell1978}, the details of the mechanism are still a matter of discussion. 
The question of the maximum energy attainable is central, especially in the context of the hunt for the sources of Galactic cosmic rays (CRs). These sources are expected to accelerate particles up to at least the PeV range and must therefore act as hadronic PeVatrons—i.e. capable of accelerating protons up to energies of order $10^{15}$ eV during at least part of their evolution \citep{drury2012,blasi2013,blasi2019,gabici2019}. Since supernova remnants have long been proposed as the dominant contributors to Galactic CRs up to the knee, the detection of an SNR operating as a PeVatron would provide compelling support for the SNR paradigm \citep{cristofari_review}.

The accelerated particles interact with the surrounding interstellar medium (ISM) to produce gamma rays. At SNR shocks, the production of VHE gamma rays is usually dominated by two mechanisms: 1) the interaction of accelerated hadrons (protons) with the ISM nuclei via the production/decay of neutral pions; 2) the interaction of accelerated electrons with soft photons (cosmic microwave background, optical, Infrared) through inverse Compton scattering. In the context of the search for hadronic PeVatrons, it is good to keep in mind that protons of energy $E_{\rm k}$ lead to gamma rays of energy $E_{\gamma} \approx E_{\rm k}/10$ and thus that 100 TeV gamma rays probe PeV protons~\citep{kelner2006}. The VHE gamma-ray domain is thus the domain of choice to study acceleration up to the PeV range. 

Remarkably, extensive studies of all known SNR shells in the VHE domain with Imaging Atmospheric Cherenkov Telescopes (IACTs), such as H.E.S.S.~\citep{HESS_SNR2018}, MAGIC~\citep{2016APh....72...76A} or VERITAS~\citep{WEEKES2002221}, have all revealed suppression in the gamma-ray spectra above a few tens of TeV, thus suggesting that the vast majority of known Galactic SNRs are not efficient PeV accelerators. 
One possible explanation for this would be that only rare SNRs from peculiar SNe, or only very young SNRs are indeed PeVatrons, and thus, we fail to detect the active PeVatron phase of an SNR~\citep{cristofari2020}.

However, G106.3+2.7 stands appart and has been claimed to be a hadronic PeVatron, especially, since high--energy gamma-rays from the region of G106.3+2.7 have been detected with a hard spectrum up to the $\sim 600$ TeV~\citep{HAWC_G106,cao2021,cao2023}. 
G106.3+2.7 is a complex source, described as a composite SNR that has been observed in radio and exhibits two regions, the head and the tail, with the head hosting a highly energetic pulsar and the associated Pulsar Wind Nebula  (PWN) called the Boomerang.
Global investigations on the multi--wavelength modeling of G106.3+2.7 - especially including gamma--ray observations with Fermi-LAT have concluded on the  presence of accelerated hadrons, reinforcing the idea that it is a hadronic PeVatron~\citep{Xin_2019,fang2022}. Moreover, the analysis of X-ray emission as a function of the distance to the pulsar has also given indications that the tail could be a hadronic PeVatron~\citep{ge2021}. 
Recent results from MAGIC have confirmed, with unprecedented angular resolution, that the VHE emission component at energies E$\gtrsim$ 10 TeV is located approximately 0.3 degrees away from the pulsar J2229+6114 \cite{abe2023}. The presence of the nearby pulsar makes the analysis and interpretation of the gamma-ray data particularly  challenging~\citep{sarkar2022,pope2023}. Setting aside the contribution from J2229+6114, we investigate the possible role of an SNR shock in producing the observed gamma-ray emission.

In Sec.~\ref{sec:particles}, we present the simple, physically motivated description of the SNR-accelerated particles used in this work. In Sec.~\ref{sec:origin}, we discuss how the MAGIC observations of the head and tail can be interpreted, and how an SNR shock can account for the entire gamma-ray spectrum detected from G106.3+2.7. Finally, in Sec.~\ref{sec:conclusions}, we summarize our conclusions.

\section{Particle acceleration and gamma rays from an SNR shock wave}
\label{sec:particles}

The content of particles accelerated at an SNR shock wave is usually described assuming that a fraction $\xi$ of the shock ram pressure $\rho u_{\rm sh}^2$ is converted into CRs, where $u_{\rm sh}$ is the shock velocity and $\rho=\mu m n_0$ is the mass density upstream. $m$ is the proton mass,  $\mu \approx 1.4$ for neutral atomic ISM, and $n_0$ the number density of hydrogen. 
If the spectrum at the shock at a given time $t$ is of the form $f(p,t)= A(t) (p/p_0)^{-\alpha}$, the CR pressure can be written $P_{\rm CR}= 4 \pi/3 \int \text{d}p p^2 f(p) p v(p) = \xi \rho u_{\rm sh}^2$, leading to $A(t) =
\frac{3\,\xi\,\rho\,u_{\rm sh}^2}
{4\pi\, p_0^{\alpha}  \int_{p_{\min}}^{p_{\max}(t)} 
\mathrm{d}p\, v(p)\, p^{3-\alpha}}$, where $v$ is the particle velocity, $p_{\rm min}$ and $p_{\rm max}$ correspond to the momentum of particles injected in the acceleration process and to the maximum momentum reached by particles, respectively. $\alpha$ is the slope of accelerated particles at the shock ($\alpha=4$ for strong shocks of compression factor $\sigma=4$ in the test-particle limit)~\citep{blasi2013}. 
In our discussion, $p_{\rm max}$ is especially important regarding the potential PeVatron nature of the SNR. At SNR shocks, the maximum energy of accelerated particles depends on the level of amplification of the magnetic field. Several works have indicated that the amplification of the magnetic field is likely due to the streaming of the shock--accelerated particles that excite instabilities in the plasma (upstream). In the usual condition met around young SNR shocks, the fastest growing instabilities are the non--resonant (Bell)~\citep{bell2004}. The maximum energy reached by particles is attained when the growth of the instabilities saturates.This typically leads to a maximum momentum for accelerated particles~\citep{bell2013,schure2013,schure2014}: 

\begin{equation}
\label{eq:pmax}
p_{\rm max}(t) \approx \frac{r_{\rm sh}(t)}{5} \frac{\xi e \sqrt{4 \pi \rho(t)} }{\Lambda} \left(\frac{u_{\rm sh}(t)}{c}\right)^{2},
\end{equation}
where $r_{\rm sh}$ is the shock radius, $\rho (t)$ is the density upstream the shock at a time $t$, and  $\Lambda=\ln\left(\frac{p_{\rm max}(t)}{mc}\right)$. 

\subsection{Particle content at the SNR shell}

The particles accelerated at the shock are trapped in the SNR shock where they are confined and can suffer adiabatic losses. The number of particles trapped in the SNR shell can simply be described via: 
\begin{equation}
\frac{\partial N(p, t) }{\partial t} = \frac{1}{p^2}\frac{\partial }{\partial p} \left[\dot{p} p^2 N(p, t)\right] + Q(p, t)
\label{eq:number}
\end{equation}
where the injection of particles at the shock is $Q(p, t) = 4\pi r_{\rm sh}^2(t) \frac{u_{\rm sh}(t)}{\sigma}  A(t) (p/p_0)^{-\alpha}$, and $4 \pi p^2 N(p, t) \text{d}p$ is the total number of particles of momentum $p$  inside the shell at a time $t$. 

At a given time $t$, particles reaching the maximum momentum $p_{\max}(t)$ are assumed to escape from the acceleration region, consistently with the standard picture of energy--dependent escape from SNR shocks~\citep{caprioli2009}. In this framework, $p_{\max}(t)$ defines the instantaneous escape boundary of the system: particles with $p < p_{\max}(t)$ remain confined within the shock region and are described by Eq.~\eqref{eq:number}, while particles reaching $p_{\max}(t)$ are released upstream and propagate into the surrounding medium~\citep{cristofari2021b}. This time--dependent escape process is explicitly taken into account in the cloud scenario discussed in Sec.~\ref{sec:cloud}, where escaping particles diffusively propagate toward dense regions and produce gamma--ray emission. In contrast, in the clump scenario (Sec.~\ref{sec:clumps}), we only consider dense clumps located within the shocked shell itself.

This description is convenient for our treatment of hadrons, for which the radiative losses (synchrotron, inverse Compton) are negligible. Inside the SNR, the velocity profile is assumed to scale as: $u(r,t)= (1-1/\sigma) u_{\rm sh}(t) r/r_{\rm sh}(t)$~\citep{ostriker1988}. This leads to a density profile steeply decreasing when moving away from the shock. Thus proton-proton interactions are only expected to be important close to the SNR shell, which motivates to simply consider a shell of width $\Delta r$ at $r_{\rm sh}(t)$ as hadronic gamma--ray emitting region. 

We adopt a fixed shell thickness $\Delta r = 0.05\,r_{\rm sh}$, which ensures the validity of the thin--shell approximation. The precise value of $\Delta r$ is not critical for our results, since variations in the effective interaction volume can be compensated by corresponding adjustments of the target density, or of the acceleration efficiency in order to reproduce the observed VHE gamma--ray normalization.

 The expected gamma rays from the energy distribution of the parent particles can be calculated using usual parametrization of the cross sections~\citep{kelner2008,khangulyan2014}. 
In addition to gamma-ray emission produced by a shock propagating in a uniform medium, a shock expanding into a clumpy medium, or the presence of a nearby molecular cloud, can significantly modify the gamma-ray emission.

\subsection{Shock expanding in a clumpy medium}
\label{sec:clumps}
The potential presence of gas clumps is of special interest for protons, as it has been shown that clumps in the ISM can substantially affect the gamma--ray emission~\citep{zirakashvili2010}. Such a clumpy medium is especially expected if the star/SNR is located in a dense molecular cloud region~\citep{denoyer1979,chevalier1999,sano2019,sano2020,schneider2022}, where some dense clumps can survive the shock wave passage, and account for most of the target material responsible for the gamma--ray emission. This possibility has been investigated in the case of RXJ1713-3946~\citep{gabici2014}. 

Let us consider clumps of typical size $L_{\rm c}= 0.1$ pc and density $n_{\rm c} \gtrsim 10^3$ cm$^{-3}$~\citep{inoue2012,fukui2012,sano2013}. As discussed in~\citet{gabici2014}, the gradient of density between the clumps and the ISM impacts the survival of clumps at the passage of the SNR shock. Defining $\chi = n_{\rm c}/n_0$,  the cloud crushing time can be defined $\tau_{\rm cc}=L_{\rm c} \chi^{1/2}/u_{\rm sh} \approx \tau_{\rm cc}   \approx 3 \times 10^{3} \left( \frac{L_{\rm c}}{0.1 \text{pc}} \right) \left( \frac{\chi}{10^4} \right)^{1/2} \left(\frac{u_{\rm sh}}{3000 \text{km/s}} \right)^{-1} \text{yr}
$  as the time it takes for the clump to be shocked~\citep{klein1994}. 
Rayleigh-Taylor and Kelvin-Helmholtz instability  typically develop on comparable timescale~\citep{klein1994}; thus $\tau_{\rm cc}$ can be used as an estimate of the typical survival time for the clumps. The survival time of the clouds is long enough so that as the shock passes, the difference of velocity between the clumps velocity and the shocked ISM drives the amplification of the magnetic field in a thin layer around the clumps. In this layer, the magnetic field can reach values greater than a few  100 $\mu$G  in timescales of tens of years~\citep{inoue2012}. Particles accelerated at the SNR shock diffuse in the ISM, and to penetrate the clump, have to go through  this thin layer around the clump $L_{\rm tr}$, assumed to be typically of size $\sim 0.05$ pc. Assuming that the ISM is sufficiently turbulent, the diffusion regime is typically the Bohm regime in which the diffusion coefficient $D_{\rm B} \approx 1/3 r_{\rm L} v$, where $r_L$ is the Larmor radius of particles. The  time needed for CRs to penetrate into the clumps diffusively is typically $\tau_{\rm cl} \approx L^2_{\rm tr}/6D_{\rm B}  \approx 4.2 \times 10^{2} \left( \frac{L_{\rm tr}}{0.05 \text{pc}} \right)^2 \left( \frac{B_{\rm tr}}{100 \mu\text{G}}\right) \left( \frac{p}{1 \text{GeV/c}}\right)^{-2} \text{yr}
$ for $p\gtrsim 1$ GeV/c.  

The total number of CRs inside the clumps is regulated by: 
\begin{equation}
\frac{\partial N_{\rm cl} (p, t)}{\partial t} = \frac{(V_{\rm cl}/V_{\rm sh})N (p, t) - N_{\rm cl}(p,t) }{\tau_{\rm cl}}
\end{equation}
where the volume of clumps is $V_{\rm cl}=4\pi/3 L_{\rm cl}^{3}$ and the volume of the SNR shell is $V_{\rm sh}$. The volume of interest of the SNR shell, in which the shocked clumps must be considered is typically the volume between the contact discontinuity and the shock. The total clump volume filling factor is taken equal to 0.01. For simplicity, we will assume that the shocked clumps do not lose mass and remain constant in size and content. This assumption is valid provided that we consider timescales shorter than $\tau_{\rm cc}$. 

\subsection{Particles accelerated at an SNR shock into a cloud}
\label{sec:cloud}

The presence of a CO cloud (close to the region of the tail)~\citep{kothes2001} naturally invites us to consider how accelerated particles can interact with molecular clouds to produce gamma rays. This scenario has been extensively discussed~\citet{gabici2007,gabici2009}. It has been especially shown that the density profile inside a cloud, as well as the diffusion regime, impact the gamma-ray emission from the cloud. 
In this scenario, the gamma-ray emission is produced by particles escaping from the SNR shock and diffusing to an external cloud. The cloud distance, size, and total mass are treated as free parameters within realistic ranges, leading to a degeneracy between diffusion properties and target mass. Consequently, variations in the diffusion coefficient can be compensated by adjusting the cloud parameters, so that the main conclusions are not sensitive to its precise value.

In this work, we do not discuss all these aspects, and, for illustrative purposes, we simply consider particles escaping from the SNR shock~\citep{caprioli2009,celli2019}, diffusively propagating into a region of higher density  (e.g., the CO dense region of the tail). This can for instance contribute to produce a harder spectrum in the tail, as discussed in Sec.~\ref{sec:head_tail}.

\section{Possible origin of the very--high--energy gamma--ray emission of G106.3+2.7}
\label{sec:origin}

\begin{figure}
	\includegraphics[width=0.49\textwidth]{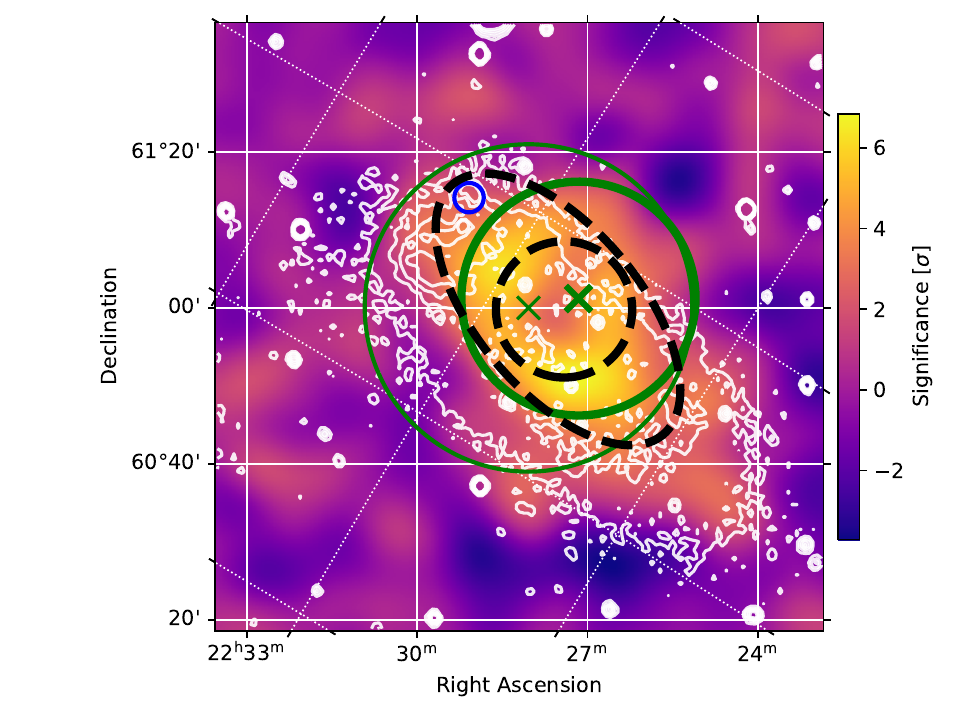}
    \caption{G106.3+2.7 as observed by MAGIC~\citep{abe2023}. The color scale shows the significance map of photons with energies above 0.2 TeV. The white contours correspond to radio observations at 1420 MHz from the Dominion Radio Astrophysical Observatory (DRAO) Synthesis Telescope~\citep{pineault2000}.
    The small solid blue circle indicates the position of J2229+6114. The green circles represent the LHAASO observations~\citep{2022A&A...658A..60Y}, with the thick circle corresponding to the WCDA data and the thin one to KM2A; the centre of the detected emission is marked with an ‘x’. The dashed black circle (diameter 0.3$^{\circ}$) and ellipse (major axis 0.9$^{\circ}$) indicate the minimal and maximal SNR shock sizes considered in this work. }
    \label{fig:Magic_1}
\end{figure}

Radio continuum measurements have revealed a head and tail structure~\citep{kothes2001}. In the GeV gamma-ray domain~\citep{fermi2009}, the emission from the head has been largely ascribed to the pulsar, the emission from tail has also been detected, and interpreted as due to accelerated hadrons~\citep{fang2022}. 
Both the head and the tail have been detected in the VHE TeV domain~\citep{abe2023} (see Fig.~\ref{fig:Magic_1}), with a spectrum from the tail harder than the one from the head.  The highest-energy gamma rays detected, up to $\sim$ 600 TeV~\citep{HAWC_G106,tibet_G106,cao2023}, clearly indicate the presence of $\sim$ PeV particles. The centroid of the $\gtrsim 100$ TeV gamma-ray emission is, moreover, located significantly away from the pulsar, although the current angular resolution of these observatories does not yet allow one to determine unambiguously whether the emission is connected to the pulsar, the head, or the tail.

\subsection{Expansion of the SNR remnant}
\label{sec:SNR}

The morphology observed in both the radio and VHE bands is elongated, and it remains unclear which regions, if any, correspond to emission produced by the SNR shock (Fig.~\ref{fig:Magic_1}).
To discuss such contribution, let us consider a SNR shock and calculate the corresponding expected gamma-ray emission. The presence of the nearby pulsar J2229+6114, if connected to the SNR, favors a core-collapse supernova (CCSN) as a progenitor. For a typical CCSN, the resulting SNR shock is expected to propagate through a circumstellar medium structured by the life of the progenitor~\citep{castor1975,weaver1977}. During its main sequence (MS), the wind blown by the massive star in the ISM of density $n_0=1$~cm$^{-3}$ inflates a hot and low density bubble of typical density $n_{\rm b}= 0.01 (n_0^{19} (u_{\rm w,2000}^2 \dot{M}_{-6})^6 t_6^{-22})^{1/35}$ cm$^{-3}$, temperature $T_{\rm b} \approx 1.6 \; 10^6 (n_0^2 (u_{\rm w,2000}^2 \dot{M}_{-6})^8 t_6^{-6})^{1/35}$ K, and size $r_{\rm b} = 28 ( (u_{\rm w,2000}^2 \dot{M}_{-6})/n_0  )^{1/5} t_6^{3/5}$ pc; where $u_{\rm w,2000}$ is the velocity of the wind in units $2000$ km/s, $\dot{M}_{-6}$ the mass-loss rate during the MS in units of $10^{-6}$ M$\odot$/yr and $t_6$ the duration of the MS in units $10^6$ yr. At the end of its life, the star enters a late sequence phase such as a Red Supergiant phase (RSG), expelling a dense low velocity wind. The gas density in the stellar wind reads $n_{\rm w, RSG}=\dot{M}_{\rm RSG}/(4 \pi m_a u_{\rm w, RSG} r^2)$, where typical values are:  $\dot{M}_{\rm RSG} \sim 10^{-5}$ M$_{\odot}$/yr for the mass-loss rate, $u_{\rm w, RSG} = 10^6$ cm/s. Equating the RSG wind ram pressure to the thermal pressure inside the bubble cavity, the radius of wind can be estimated: $r_{\rm w, RSG}=(\dot{M} u_{\rm w,RSG}/(4 \pi k_{\rm b}n_{\rm b} T_{\rm b}))^{1/2}$ where $k_{\rm b}$ is the Boltzmann constant. 
In fact, since massive stars often form in associations and clusters, one could argue that the collective effects of the stellar winds form a cavity of size substantially larger. In the case of bubbles and superbubbles inflated, cavity of sizes $R_{\rm b, cluster} \approx 270 (L_{\rm 38, tot}/n_0)^{1/5} t_{\rm 10 Myr}^{3/5}$ pc~\citep{maclow1988} with $L_{\rm 38, tot}$ the total luminosity of the winds in units $10^{38}$ erg/s and $t_{\rm 10 Myr}$ the age of the cluster/association in units 10 Myr; with a corresponding low density cavity of density $n_{\rm b, cluster} \approx 0.01 (L_{\rm 38, tot}^6 n_0^{19} t_{\rm 10 Myr}^{-22} )^{1/35} $ cm$^{-3}$. The density of the cavity in this case is typically of the same order than in the case of a single star, the main difference being only the size of the cavity. 

The SN explosion drives a spherically symmetric shock that expands through the circumstellar environment, first traversing a dense RSG wind, then a low-density cavity, and ultimately reaching the unperturbed ISM. The time evolution of the shock radius and velocity can be described analytically using the thin-shell approximation~\citep{bisnovatyikogan1995,ptuskin2005}.

\begin{figure}
	\includegraphics[width=0.45\textwidth]{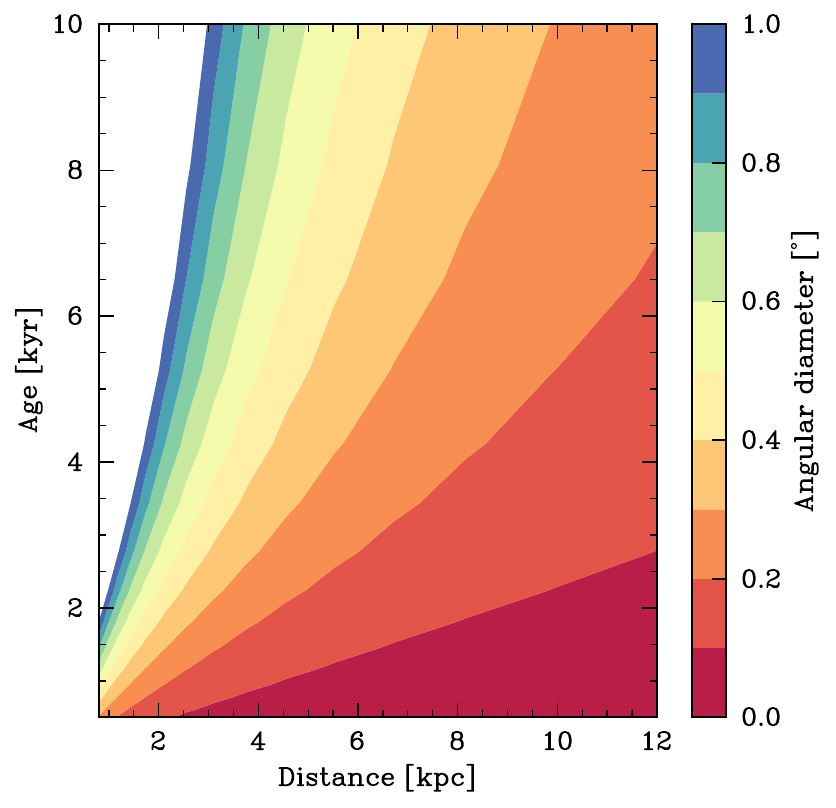}
    \caption{Angular diameter of the SNR shock (2$\times r_{\rm sh}$ as a function of its distance and age. The SNR shock expands in a structured CSM, first through the dense RSG wind and then through a low--density bubble ($n_{\rm b}=10^{-2}\,\mathrm{cm}^{-3}$), in which it evolves for most of its lifetime.
    } 
    \label{fig:contour}
\end{figure}

Both the distance and the age of G106.3+27 remain poorly constrained: with values for the distance found in the range [$\sim 800$ pc, $\sim$12 kpc]~\citep{pope2023}, and a pulsar age estimated from the spin-down rate to be at most $\approx 10$ kyr~\citep{halpern2001}. 
As illustrated in Fig.~\ref{fig:contour}, the apparent angular size can be used to constrain the pair (distance, age). We consider the evolution of the SNR diameter following a core-collapse supernova (CCSN) with typical parameters ($E_{\rm SN}=10^{51}$~erg, $\dot{M} = 10^{-5}\,{\rm M}_{\odot}\,{\rm yr}^{-1}$, and other values specified in Sec.~\ref{sec:origin}). Fig.~\ref{fig:contour} shows the resulting SNR size as a function of age and distance.
Assuming that the observed extension in the TeV range of $\sim 0.3$--$0.4^{\circ}$ corresponds to the SNR diameter (e.g., the dashed black circle in Fig.\ref{fig:Magic_1}), we find that distances $\lesssim 4$~kpc typically imply an age of $2.5$--$3.5$~kyr, whereas larger distances $\gtrsim 8$~kpc favor ages $\gtrsim 6$~kyr. Conversely, if the SNR diameter were significantly larger of $\sim 0.6$--$0.8^{\circ}$ (e.g. the major axis of the dashed black ellipse in Fig.\ref{fig:Magic_1}), then distances $\lesssim 4$~kpc are preferred for the entire age range considered.

We emphasize that the elongated morphology of the source shown in Fig.~\ref{fig:Magic_1} is not used as a geometrical input in the modeling. We focus on the simplified situation of spherical SNR shock framework to discuss the typical apparent extension scale and infer plausible age--distance ranges. The observed elongation is therefore treated as a qualitative morphological feature (e.g. due to projection effects, of large--scale density gradients in the ambient medium), without direct impact on the spectral modeling. The TeV spectral interpretation with a single SNR shock remains fully compatible with the observed elongated morphology. In particular, a single SNR scenario can naturally account for both the head and tail emission components, and none of the acceptable spectral solutions are disfavored on spatial grounds such as centroid location or source extension.

\begin{figure}
	\includegraphics[width=0.49\textwidth]{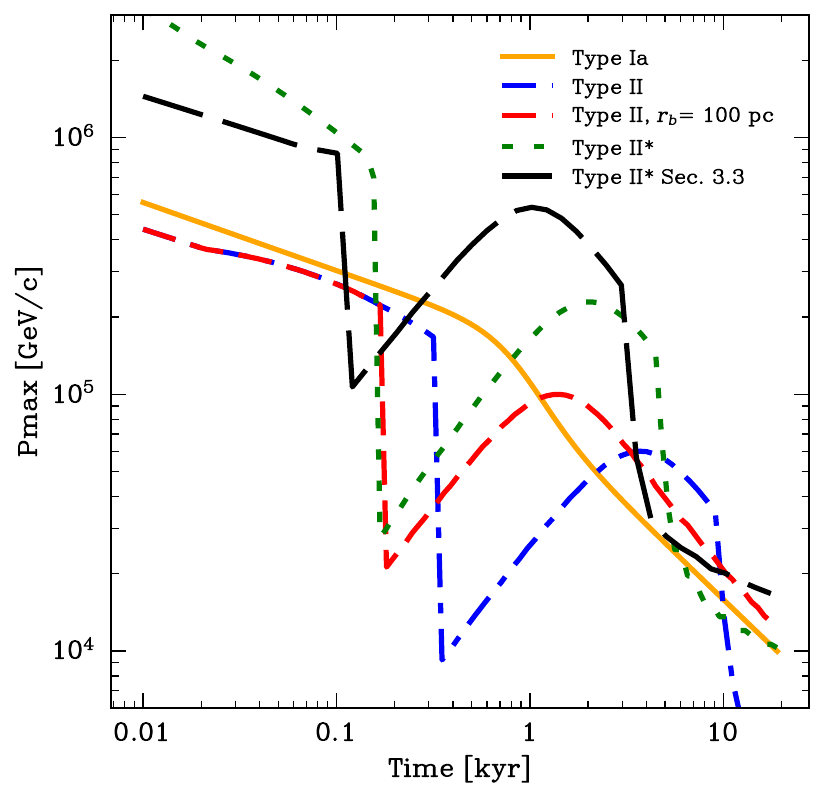}
    \caption{Time evolution of the maximum momentum of accelerated particles when the amplified magnetic field is driven by the growth of non--resonant streaming instabilities~\citep{bell2013}. Type Ia (solid yellow), Type II (dot-dashed blue), Type II with an extended low density bubble of size $r_{\rm b}=100$ pc (dashed red), and Type II* (dotted green) SN progenitors are considered. The best-fit found in Sec.~\ref{sec:broad} is shown in dashed black.}
    \label{fig:pmax}
\end{figure}

\subsection{The TeV emission from the head and tail}
\label{sec:head_tail}

As discussed in Sec.~\ref{sec:particles}, the maximum energy at SNR shocks is conditioned by the growth of non--resonant instabilities. 
For typical SNRs from  from CCSNe (Type II), the maximum energy remains below $\lesssim 100$ TeV for ages $\gtrsim 1$ kyr,   considering bubble of typical radius $r_{\rm b}\approx 28$ pc, or an extended bubble radius of 100 pc.
The same also applies for typical SNRs for thermonuclear (Type Ia) supernova, whose progenitor typically expands into a low-density ambient medium, the maximum energy remains below $\lesssim 100$ TeV for ages $\gtrsim 1$ kyr (Fig.~\ref{fig:pmax}).
Only unusual CCSNe (type II*) with high total explosion energy $E_{\rm SN} \gtrsim 5 \; 10^{51}$ erg, high mass--loss rates $\dot{M} \gtrsim 10^{-4}$ M$_{\odot}$/yr and ejecta mass $M_{\rm ej} \lesssim 2$ M$_{\odot}$ can attain $\sim 100$ TeV at $\sim 7-8$ kyr while the SNR shock is expanding in the low-density main-sequence inflated bubble. 
This means, that if we rely on usual parameters for the SNR evolution, the detected 10 TeV gamma rays seem to root in favor of a young, of typically $\sim 1$ kyr old SNR. Coupled with the arguments on the angular extension, this would thus favor a distance $\lesssim 2-3$ kpc.

\begin{figure}
	\includegraphics[width=0.49\textwidth]{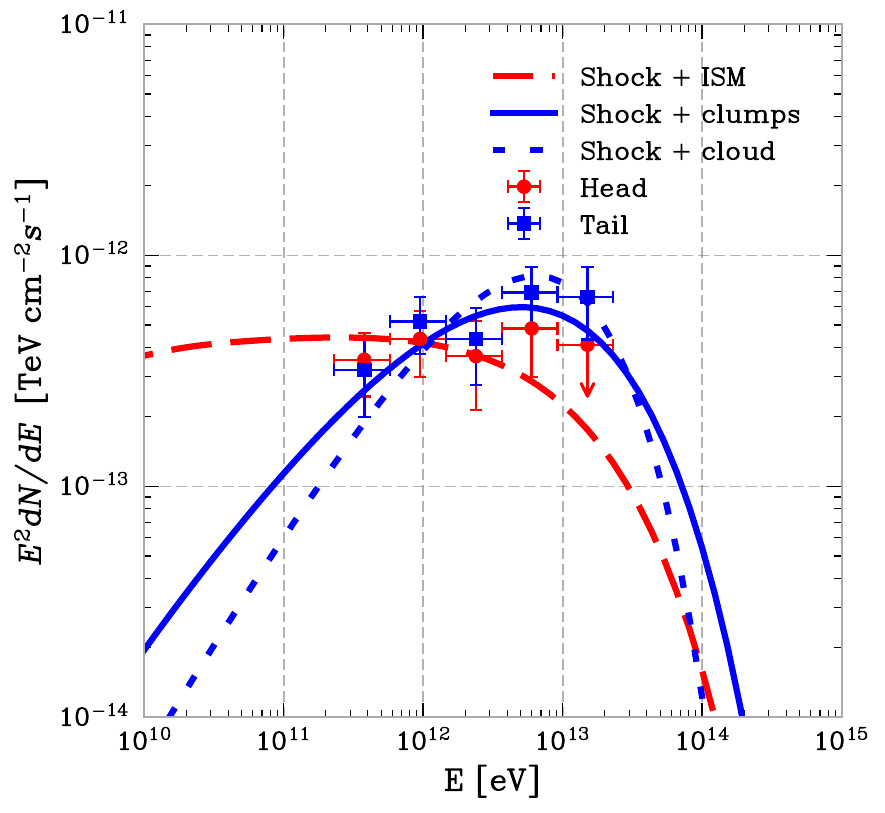}
    \caption{Gamma-ray emission from a SNR shock from a Type II SNe, expanding in a uniform medium (dashed red), propagating into a clumpy medium (solid blue), and from protons that have escaped the shock and interact with a cloud of density $10^2$ cm$^{-3}$, radius 5 pc, located at 10 pc from the SNR shock wave (dotted blue). The first scenario readily accounts for the MAGIC observations of the head (red data points), whereas the second and third scenarios can explain the MAGIC observations of the tail (blue data points).  
    }
    \label{fig:Spectrum1}
\end{figure}

The ability of an SNR shock to account for the TeV emission from both the head and the tail is illustrated in Fig.~\ref{fig:Spectrum1}, assuming a SNR from a CCSNe as described in the previous Section, expanding for 1 kyr at a distance of 1 kpc. The age of the SNR is adopted for illustrative purposes and longer ages (larger distances) also allow for a decent fit to the TeV data.  Several scenarios can explain the harder spectrum observed in the tail by MAGIC, and recently confirmed by the LST~\citep{LST_proceeding,Carrasco:20251O}. We rely on the situations described in Sec.~\ref{sec:particles} and show that the MAGIC spectrum of the head can be reproduced by an SNR shock expanding in the ISM, whereas the harder spectrum of the tail can be explained either by a shock propagating into clumps or by particles escaping from the SNR shock into denser regions. The cloud and clumps scenarios are especially supported by the observations of a CO cloud associated with the tail, implying a higher proton density in the tail region~\citep{kothes2001}. In such a dense molecular environment, as discussed in Sec.~\ref{sec:particles}, the presence of clumps is expected and can significantly affect the gamma-ray emission~\citep{zirakashvili2010}.
IRAM 30 m CO observations toward G106.3+2.7 confirm the presence of molecular gas coincident with the gamma-ray–emitting tail, but do not reveal unambiguous signatures of a direct shock–cloud interaction, suggesting that the cloud may instead be illuminated by escaping CRs~\citep{liu2022}. The distance to the CO cloud has been estimated to be $\sim 0.8$ kpc, and thus assuming that the gamma-ray emission is associated with the cloud implies that the age of the SNR must be $\lesssim 1$ kyr (Fig.~\ref{fig:contour}). As mentioned previously, such values also allow to satisfactorily account for the TeV observations. The diffusion coefficient adopted for illustrative purposes is $D\approx 0.05 \times 10^{28} (E/10 \; \text{GeV})^{0.5}$~\citep{gabici2009}, but the actual value does not significantly affect our conclusions.

Let us additionally mention that a SNR leptonic scenario could also account for the spectra measured in the tail and head, as for instance discussed in~\citet{abe2023}. The difference in the spectral indices can be explained by assuming different maximum energies for the electrons in this two zones. The maximum energy for electrons is usually set by the losses competing with the acceleration process. A different diffusion regime and average magnetic field in the head and tail could naturally help produce the spectrum observed by MAGIC.
In both hadronic and leptonic scenarios, a young typical SNR shock of age $\sim 1$ kyr would correspond to a maximum energy of accelerated particles remaining below $\sim 100$ TeV and is not expected to account for the acceleration of $\sim$ PeV particles seen by HAWC, AS$\gamma$ Tibet and LHAASO. These particles would thus have to be connected to the pulsar/pulsar wind, or to a past activity of PeV accelerating SNR shock. In the next section, we will investigate the properties required for a SNR to account for the entire gamma-ray emission, including the observations at the highest energies. 

In both hadronic and leptonic scenarios, a young and otherwise typical SNR shock of age $\sim 1$~kyr is not expected to accelerate particles beyond $\sim 100$~TeV, and therefore cannot account for the $\gtrsim 100$~TeV gamma--ray emission observed by HAWC, Tibet AS$\gamma$, and LHAASO. The highest--energy particles would thus have to be associated either with the pulsar and its pulsar wind nebula (e.g. via PeV electron or hadron acceleration in pulsar--driven outflows \citep{aharonian2012,fang2022,sarkar2022,wilhelmi2022}, or with a past PeVatron phase of the SNR shock. In this latter case, ``past activity'' refers to an early evolutionary phase of the SNR ($\lesssim$ few $10^2$~yr), during which PeV energies could be reached, with the highest--energy particles having subsequently escaped the shock and propagated into the surrounding medium~\citep{gabici2009,blasi2025}.

\subsection{Accounting for the broadband Gamma-Ray Spectrum with a Single SNR}
\label{sec:broad}

We investigate whether the full gamma-ray emission (from the GeV to several hundred TeV) spectrum from G106.3+2.7 can be explained by a single SNR and explore the parameter space that yields the best-fit spectrum from GeV to multi-TeV energies.
We perform a systematic exploration of the  parameter space of the SNR model described in Sec.~\ref{sec:particles}, considering a SNR from a CCSNe (Type II), and considering only the gamma-ray emission from accelerated protons. We thus aim to identify the models that best reproduce the observations from Fermi-LAT~\citep{2019ApJ...885..162X}, VERITAS~\citep{2009ApJ...703L...6A}, and LHAASO~\citep{2022A&A...658A..60Y}. In the TeV range, we use the VERITAS data instead of MAGIC, as they cover the same energy range and appear slightly more consistent with the Fermi-LAT data. We considered 7 parameters: $\dot{M}$, $E_{\rm SN}$, $M_{\rm ej}$,  $\xi$, $\alpha$, the age of the supernova, and the distance. 
The parameter space exploration was done in a uniform grid described in Tab.~\ref{tab:grid}.

\begin{table*}[]
    \centering
    \begin{tabular}{c|c|c|c|c|c|c}
    Parameter & min & max & step& best fit & most recurring & preferred range \\
    \hline
    $\log_{10}(\dot{M})$ [M$_{\odot}$/yr] & -5 & 0 & 1 & -5 & -1 & -5$-$0 \\
    $E_{\rm SN}$ [$10^{51}$ erg] & 0.5 & 10.0 & 0.5 & 9.5 & 10.0 & 7.0$-$10.0 \\
    $M_{\rm ej}$ [$M_{\odot}$] & 1 & 10 & 1& 8& 2 & 1$-$7\\
    $\xi$ & 0.01 & 0.31 & 0.03& 0.25 & 0.31 & 0.22$-$0.31\\
    $\alpha$ & 3.5 & 4.5 & 0.1& 4.4 & 4.5 & 3.5$-$4.5 \\
    $\text{age} \text{ [kyr]}$ & $0.626$ & $10$ & $\times 1.41$& $0.886$ & $0.886$ & $0.626$$-$$1.252$\\
    distance [kpc]& 0.8 & 10.8 & 0.524 & 3.9 & 5.0 & 2.9$-$6.0 \\
    \end{tabular}
    \caption{Parameter space considered  for the modeling of a CCSNe SNR and properties of the models best fitting to the broadband gamma-ray emission (see Sec.~\ref{sec:broad}). The best fit is column gives the parameters of the single best model. Most recurring is the parameter value appearing in most of the selected models. Preferred range gives the range of parameters appearing in at least 50\% as many models as the most recurring one.}
    \label{tab:grid}
\end{table*}

\begin{figure}
	\includegraphics[width=0.49\textwidth]{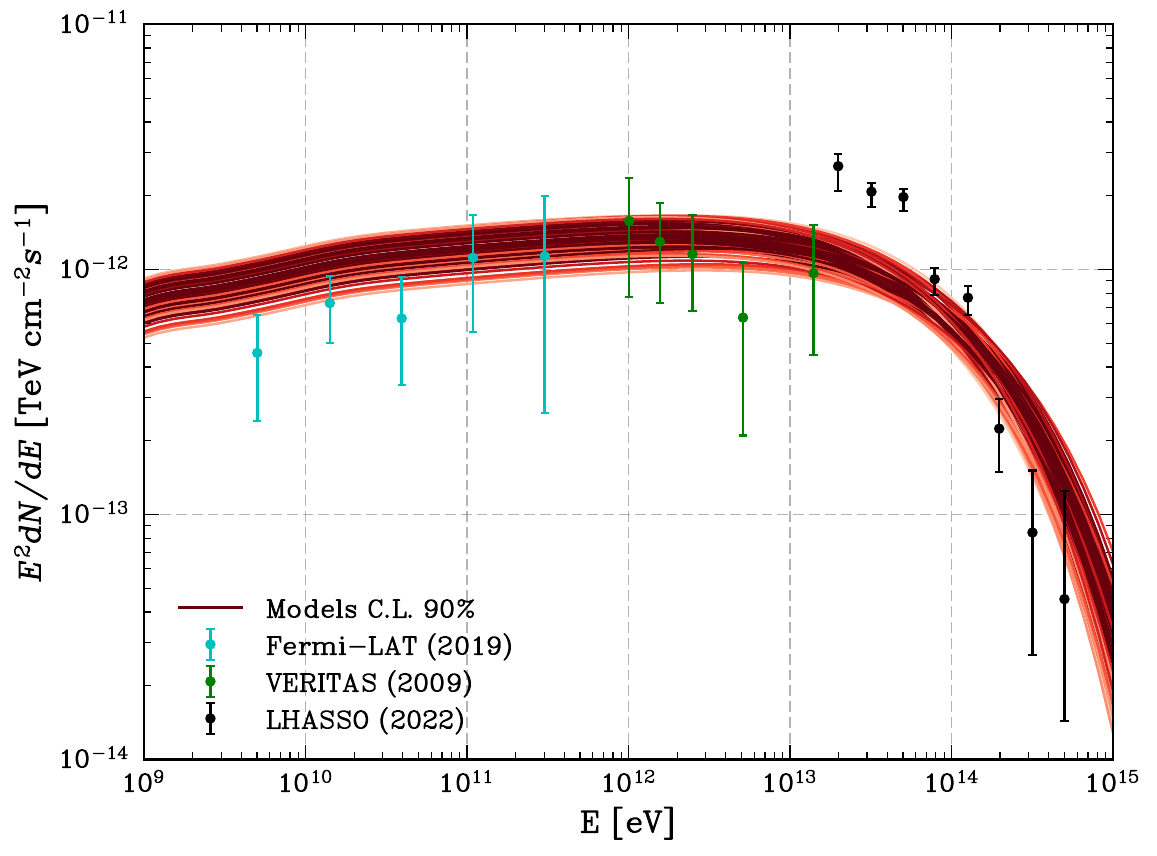}
    \caption{Best-fitting hadronic models obtained from our grid search. Flux points from each instrument were used to select the models. Observationnal data points are taken from Fermi-LAT~\citep{2019ApJ...885..162X}, VERITAS~\citep{2009ApJ...703L...6A}, and LHAASO~\citep{2022A&A...658A..60Y}. The color of the models' curves go from dark to light-red going from the best to worse model likelihood.}
    \label{fig:gridmod}
\end{figure}

To select the best solution, we compute the model log-likelihood, assuming a Gaussian likelihood for each of the previously mentioned spectral points and neglecting instrumental systematics. Using $f_{\rm model}$ for the model-predicted flux, $f_{\rm data}$ for the measured flux, and $\sigma_{\rm data}$ for the associated uncertainty, the log-likelihood is given by:
\begin{equation}
    -2 \ln\mathcal{L} = \sum \frac{(f_{\rm model} -f_{\rm data})^2}{\sigma_{\rm data}^2} 
\end{equation}
The quantity $\Delta \mathcal{L} = (-2\ln\mathcal{L} - \min(-2\ln\mathcal{L}))$ behaves as a $\chi^2$ distribution with 7 degree of freedom. We can thus select solutions with $\Delta \mathcal{L}$ of less than 12.017 to keep models in the 90\% C.L. region. The selected models are shown in Fig.~\ref{fig:gridmod} together with the spectral points. We do not include in our uncertainties the systematics arising from different integration regions and between instruments.

Tab.~\ref{tab:grid} summarizes the explored parameter space and the properties of the models retained within the 90\% confidence level. We find that the preferred solutions favor a young SNR, with an age $\simeq 0.6$--$1.3$~kyr, located at intermediate distances of $\sim 3$--$6$~kpc, and characterized by large explosion energies (typically $E_{\rm SN} \gtrsim 7 \times 10^{51}$~erg) and high acceleration efficiencies ($\xi \gtrsim 0.22$). The best-fitting models tend to lie near the upper boundary of the explored $E_{\rm SN}$ and $\xi$ ranges, indicating that extreme energetics are required for a single SNR scenario to reproduce the broadband gamma-ray data. While $M_{\rm ej}$ remains only weakly constrained, the number of acceptable models decreases toward large $M_{\rm ej}$, favoring moderate values. 
In contrast, the mass-loss rate $\dot{M}$ spans almost the full explored range, showing no strong statistical preference, which indicates a significant degeneracy between the CSM structure and other dynamical parameters.
Overall, reproducing the GeV to multi-TeV emission with a single SNR requires both a young system and an unusually efficient conversion of shock energy into CRs.

The favored momentum spectral index of accelerated particles at the shock is around $\alpha \simeq 4.5$, softer than the canonical test-particle DSA prediction $\alpha = 4$ for strong unmodified shocks. Such soft spectra can naturally arise if the effective compression ratio is reduced, for instance in moderately weak shocks propagating in a hot stellar-wind bubble, or when Afvenic drift in amplified upstream magnetic fields decreases the velocity jump experienced by the particles~\citep{zirakashvili2008b,haggerty2020,cristofari2022}. At the same time, acceptable solutions are also found for all considered values of $\alpha$, including for harder spectra ($\alpha \lesssim 4$). Hard spectral indices can be produced in non-linear DSA, where the back-reaction of accelerated particles modifies the shock structure and leads to concave spectra that become harder than $\alpha = 4$ at high energies \citep{mckenzie1982,malkov2001,amato2005}. Hard effective indices may also arise in radiative or highly compressed environments, where cooling and enhanced compression alter the acceleration conditions \citep{metzger2015,cristofari2025}. Overall, all these values remain somewhat compatible with DSA operating in realistic SNR environments.
We note that acceptable fits are obtained across the full explored range of $\alpha$ ($3.5$--$4.5$), with no strong statistical preference, so that the reported best-fit value reflects only a mild preference within a largely degenerate parameter space.

\section{Conclusions}
\label{sec:conclusions}
Several lines of evidence suggest that the gamma-ray emission from G106.3+2.7 originates from an SNR shock. However, the connection of the tail and head regions to this shock, the age and distance of the SNR, its possible association with the pulsar J2229+6114, and its ability to accelerate particles to PeV energies remain uncertain.

Relying on a simple analytical modeling describing the time evolution of the shock dynamics and particle acceleration at a strong shock, we discuss the possibility of accounting for the TeV gamma-ray emission with a single SNR shock, and for the entire gamma-ray spectrum from the GeV to the $\sim 100$ TeV range. 
Our conclusions are the following:

1) If the emission of the head and tail are both powered by the SNR shock, the gamma--ray spectra extending up to the $\sim 10$ TeV range favor a young SNR (of typical age $\lesssim 1$ kyr), accelerating $\sim 100$ TeV particles. 

2) The harder spectrum in the tail can be accounted with the presence of a clumpy region or the presence of a molecular cloud, as suggested by CO data. The connection between the tail and the CO cloud suggests a short distance $\sim 0.8$ kpc to the object G106.3+2.7, that would thus favor a young age $\lesssim 1$ kyr to account for the observed angular extension.

3) An SNR with "usual" physical parameters is not expected to copiously accelerate PeV particles. Under the consensual picture of DSA where the maximum energy is set by growth of non-resonant modes excited by the streaming of CRs, unusually rare SNRs are needed. Although this is possible, such SNRs are expected to be rare ($\lesssim 5$ \% of all SNe)~\citep{cristofari2020}. To explain the $\sim 100$ TeV emission (due to $\sim$ PeV CRs), this would either mean that we are observing a very unusual SNR, or some  important element of our understanding of particle acceleration at SNR shocks is missing.

This favors the idea that the $\sim 100$ TeV emission is due to particles accelerated at the pulsar/ pulsar wind rather than at the SNR shock wave, and that G106.3+2.7 is not a hadronic SNR PeVatron, in the sense that the PeV particles are not directly due DSA at the SNR shock wave.  

An alternative and physically well-motivated interpretation is provided by hybrid SNR-pulsar scenarios, in which the SNR shock dominates the gamma-ray emission up to the multi-TeV range ($\sim 10$--$30$~TeV), while the pulsar and/or its pulsar wind nebula (PWN) becomes the dominant contributor at higher energies. Within such a framework, standard SNR parameters are sufficient to account for the GeV-TeV emission, whereas the $\gtrsim 100$~TeV component is attributed to particle acceleration in pulsar-driven outflows or nebular structures. This hybrid picture alleviates the need to invoke unusually high explosion energies, mass-loss rates, or acceleration efficiencies in the SNR component, and provides a coherent interpretation of the broadband spectrum. More generally, composite SNR-pulsar scenarios have been proposed for G106.3+2.7 and similar systems \citep{Xin_2019, abe2023}.

The case of G106.3+2.7 illustrates the challenges inherent in interpreting VHE $\gamma$-ray emission from complex environments, where not only individual objects, such as the SNR shock, or the pulsar, but also their environment shall be taken into account~\citep{martin2024}. In particular, the presence of $\gtrsim 100$~TeV $\gamma$ rays must be assessed with caution, as misinterpretation may lead to erroneously identifying an astrophysical object as a PeVatron. Current evidence indicates that SNR shocks-even those claimed to be associated with unusual CCSNe, such as in Cas~A-are generally inefficient at accelerating particles to PeV energies \citep{blasi2025}. In the coming years, many regions hosting expanding SNR shocks will likely be detected with PeVatron-like signatures; however, this does not necessarily imply that the SNR shocks themselves are the responsible PeV accelerators \citep{IC443_LHAASO}.
Future observations with increased angular resolution in the $\gtrsim 10 -100$ TeV range such as with CTAO will be essential to study the morphology of the source in the VHE gamma--ray domain, and understand the site responsible of the acceleration of PeV particles~\citep{CTA_PeVatron}.

\begin{acknowledgements}
PC acknowledges support from the GALAPAGOS PSL Starting Grant. PC and HC acknowledge support from the GATTACA project funded by Action Thématique Phénomènes Extrêmes et Multi-messagers (ATPEM). GE acknowledges financial support from the Severo Ochoa grant CEX2021-001131-S funded by MCIN/AEI/ 10.13039/501100011033.
\end{acknowledgements}

\bibliographystyle{aa} 
\bibliography{G106.bib} 

\end{document}